\documentclass[prb,preprint]{revtex4-2}

\usepackage{amsmath}  
\usepackage{amsfonts} 
\usepackage{graphicx} 

\begin{document}

\title{Transformation Between the Schwarzschild Coordinates and Local
  Inertial Coordinates}

\author{Frank Wang} \email{fwang@lagcc.cuny.edu}
\affiliation{Department of Mathematics, LaGuardia Community College of
  the City University of New York, Long Island City, NY 11101}

\date{\today}

\begin{abstract}
  We present a transformation between the Schwarzschild coordinates
  and local inertial coordinates, and demonstrate the effect of
  gravitational bending of light near a massive body in a small
  region.  When a photon is emitted from a point near the Earth's
  surface with an initial horizontal direction, its parabolic
  trajectory would have a vertical deflection that is about three
  times that of a non-relativistic Newtonian particle would trace.
\end{abstract}

\maketitle 

In 2024, Balbus claimed that the common equivalence prince explanation
of gravitational bending of light is quantitatively incorrect by a
factor for three in the Schwarzschild geometry.\cite{balbus} Typical
books assert that the vertical deflection of light in an stationary
elevator on earth is $-\frac{1}{2} g t^{2}$, where $g$ is the
acceleration due to gravity and $t$ is a photon's time of flight
across the elevator, but his calculation indicated that it should be
$-\frac{3}{2} g t^{2}$.  Wang followed up with a comment, which
corroborated with Balbus's result but with a relativistic correction
factor.\cite{wang} In 2004, D{\'i}az-Miguel also found a factor of
three in the vertical fall of terrestrial light,\cite{diaz} but Linet
disputed his interpretation of coordinates and insisted $-\frac{1}{2}
g t^{2}$ was correct.\cite{linet} Because Linet used spherical
coordinates and Wang used Cartesian coordinates, we here present
another transformation for a direct comparison.

With the commonly used symbols $G$ for the gravitational constant and
$c$ for the speed of light, the Schwarzschild metric for a mass $M$ is
\begin{equation}\label{eq:schwarzschild0}
  ds^{2} = - (1-2 r_{g}/r) c^{2} dt^{2} + (1-2 r_{g}/r)^{-1} dr^{2} +
  r^{2} (d \theta^{2} + \sin^{2} \theta d \phi^{2}),
\end{equation}
where $r_{g}=GM/c^{2}$.  Our goal is to find a coordinate system
$x^{\prime \mu}=(T, X, Y, Z)$ to make this metric locally flat near a
point $P$ in this form:
\begin{equation}\label{eq:locallflato2}
  ds^{2} = -c^{2} d T^{2} + dX^{2} + dY^{2} + dZ^{2} + O[(x^{\prime
      \alpha})^{2} ] dx^{\prime \mu} dx^{\prime \nu}.
\end{equation}
A different way to state the locally flat condition is that the first
derivative of the metric elements is zero: $\partial g_{\mu
  \nu}/\partial x^{\prime \alpha} =0$ at $P$.\cite{price}

We exploit Albert Einstein's ``happiest thought'' which occurred to
him in 1907: ``\textit{for an observer falling freely from the roof of
  a house there exists}---at least in this immediate
surroundings---\textit{no gravitational field}'' [his
  italics].\cite{pais} Steven Weinberg's interpretation of the
Equivalence Principle is the following: ``at any spacetime point in an
arbitrary gravitational field there is a `locally inertial' coordinate
system in which the effects of gravitation are absent in a
sufficiently small spacetime neighborhood of that
point.''\cite{weinberg} Imagine an observer inside a free-falling
elevator.  It is natural to use rectangular $X$, $Y$, $Z$ grid
attached to the walls as spatial coordinates, and a fixed clock on the
wall to measure time $T$.  According to Einstein's insight, the metric
written in the coordinate system $(T, X, Y, Z)$ should automatically
satisfy Eq.~(\ref{eq:locallflato2}).

Using the method described in Wang's paper,\cite{wang} but without
converting Eq.~(\ref{eq:schwarzschild0}) to Cartesian coordinates, we
found that the following transformation near a point $P$ with $t=0$,
$r=r_{0}$ ($r_{0}> 2 r_{g}$), $\theta = \pi/2$, $\phi=0$,
\begin{multline}\label{eq:sch2}
  t = \frac{T}{\sqrt{1 - \frac{2 r_{g}}{r_{0}}}}
  -\frac{r_{g}}{r_{0}^{2}} \frac{1}{\left(1 - \frac{2
      r_{g}}{r_{0}}\right)} T Z, \\ r-r_{0} = \sqrt{1 - \frac{2
      r_{g}}{r_{0}}} \, X + \frac{r_{g}}{2 r_{0}^{2}} X^{2} +
  \frac{\left(1 - \frac{2 r_{g}}{r_{0}} \right)}{2 r_{0}}(Y^{2} +
  Z^{2}) - \frac{r_{g}}{2 r_{0}^{2}} c^{2} T^{2}, \\ \theta -
  \frac{\pi}{2} = \frac{Z}{r_{0}} - \frac{\sqrt{1 - \frac{2
        r_{g}}{r_{0}}}}{r_{0}^{2}} XZ, \quad \phi = \frac{Y}{r_{0}} -
  \frac{\sqrt{1 - \frac{2 r_{g}}{r_{0}}}}{r_{0}^{2}} XY,
\end{multline}
will make Eq.~(\ref{eq:schwarzschild0}) locally flat in the form of
Eq.~(\ref{eq:locallflato2}).

Linet's transformation between the Schwarzschild metric and a small
neighborhood with uniform gravitational field (or constant
acceleration) is
\begin{multline}\label{eq:linet3}
  t = \frac{T}{\sqrt{1 - \frac{2 r_{g}}{r_{0}}}} , \quad r-r_{0} =
  \sqrt{1 - \frac{2 r_{g}}{r_{0}}} \, X + \frac{r_{g}}{2 r_{0}^{2}}
  X^{2} + \frac{\left(1 - \frac{2 r_{g}}{r_{0}} \right)}{2
    r_{0}}(Y^{2} + Z^{2}), \\ \theta - \frac{\pi}{2} = \frac{Z}{r_{0}}
  - \frac{\sqrt{1 - \frac{2 r_{g}}{r_{0}}}}{r_{0}^{2}} XZ, \quad \phi
  = \frac{Y}{r_{0}} - \frac{\sqrt{1 - \frac{2
        r_{g}}{r_{0}}}}{r_{0}^{2}} XY,
\end{multline}
in which we modified his notation to make it consistent with ours.
The resulting metric is
\begin{equation}\label{eq:linet2}
  ds^{2} = -(1 + 2 g_{\mathrm{S}} X) \, dT^{2} + dX^{2} + dY^{2} +
  dZ^{2}.
\end{equation}
where the Schwarzschild acceleration $g_{\mathrm{S}}$ is defined as
\begin{equation}\label{eq:gSch}
  g_{\mathrm{S}} = \frac{c^{2} r_{g}}{r_{0}^{2} \sqrt{1- \frac{2
        r_{g}}{r_{0}}}} = \frac{1}{\sqrt{1- \frac{2 GM}{c^{2} r_{0}}}}
  \frac{GM}{r_{0}^{2}}.
\end{equation}

Because a free-falling frame, not a constant-acceleration one, plays a
central role in general relativity, our transformation,
Eq.~(\ref{eq:sch2}), is more suitable for describing geodesics.
Consider a photon emitted at $X=0$ and $Z=0$ in the $XY$-plane in the
$Y$ direction, after a flight the position is
\begin{equation}
  X = Z = 0, \quad Y = cT.
\end{equation}
In terms of the Schwarzschild coordinates,
\begin{equation}\label{eq:rdisplacement}
  r = r_{0} + \frac{c^{2} T^{2}}{2 r_{0}} - \frac{r_{g}}{2 r_{0}^{2}}
  c^{2} T^{2} - \frac{r_{g}}{r_{0}^{2}} c^{2} T^{2} , \quad \phi =
  \frac{cT}{r_{0}}.
\end{equation}
If there is no gravitational effect, the trajectory of a photon would
be $r_{0} \sec \phi$, which can be approximated as
\begin{equation}
  r_{0} \sec\left( \frac{cT}{r_{0}} \right) \approx r_{0} +
  \frac{c^{2} T^{2}}{2 r_{0}} , \quad cT \ll r_{0}.
\end{equation}
The radial deviation is therefore $-\frac{3}{2} g T^{2}$, where
$g=c^{2} r_{g}/r_{0}^{2} = GM/r_{0}^2$.  The proper
distance\cite{wang,mtw} is
\begin{equation}\label{eq:ell}
  \ell = - \frac{3}{2} \frac{g}{\sqrt{1 - \frac{ 2r_{g}}{r_{0}}}}
  T^{2} .
\end{equation}
We examine Linet's approach: in a uniform gravitational field
described by the metric shown Eq.~(\ref{eq:linet2}), the geodesic is
\begin{equation}
  X = - \frac{1}{2} g_{\mathrm{S}} T^{2}, \quad Y = cT , \quad Z = 0.
\end{equation}
Substituting these coordinates into the $r$-coordinate in
Eq.~(\ref{eq:linet3}), dropping the $X^{2}$ term as Linet did, we also
obtain Eq.~(\ref{eq:rdisplacement}).

We conclude that when the light propagates horizontally near Earth's
surface (at the right angle to the direction of gravitational
acceleration), the vertical deflection after a horizontal displacement
of $cT$ is $-\frac{3}{2} g_{\mathrm{S}} T^{2}$, not $-\frac{1}{2}
g_{\mathrm{S}} T^{2}$, where $g_{\mathrm{S}}$ is defined in
Eq.~(\ref{eq:gSch}).  The factor of three is physical.

\end{document}